\documentclass[twocolumn,superscriptaddress,preprintnumbers,amsmath,amssymb,prd]{revtex4}
\usepackage{graphicx}
\UseRawInputEncoding

\begin{document}

\thispagestyle{empty}

\title{Quantum field theoretical framework for the electromagnetic response of
graphene and dispersion relations with implications to the Casimir effect
}

\author{
G.~L.~Klimchitskaya}
\affiliation{Central Astronomical Observatory at Pulkovo of the
Russian Academy of Sciences, Saint Petersburg,
196140, Russia}
\affiliation{Peter the Great Saint Petersburg
Polytechnic University, Saint Petersburg, 195251, Russia}
\author{
V.~M.~Mostepanenko}
\affiliation{Central Astronomical Observatory at Pulkovo of the
Russian Academy of Sciences, Saint Petersburg,
196140, Russia}
\affiliation{Peter the Great Saint Petersburg
Polytechnic University, Saint Petersburg, 195251, Russia}

\begin{abstract}
The spatially nonlocal response functions of graphene obtained on the
basis of first principles of quantum field theory using the polarization
tensor are considered in the areas of both the on-the-mass-shell and
off-the-mass-shell waves. It s shown that at zero frequency the
longitudinal permittivity of graphene is the regular function, whereas
the transverse one possesses a double pole for any nonzero wave vector.
According to our results, both the longitudinal and transverse
permittivities satisfy the dispersion (Kramers-Kronig) relations
connecting their real and imaginary parts, as well as expressing each of
these permittivities along the imaginary frequency axis via its imaginary
part. For the transverse permittivity, the form of an additional term
arising in the dispersion relations due to the presence of a double pole
is found. The form of dispersion relations is unaffected by the branch
points which arise on the real frequency axis in the presence of
spatial nonlocality. The obtained results are discussed in connection
with the well known problem of the Lifshitz theory which was found to
be in conflict with the measurement data when using the much studied
response function of metals. A possible way of attack on this problem
based on the case of graphene is suggested.
\end{abstract}

\maketitle

\section{Introduction}

The 2D-sheet of carbon atoms known as graphene \cite{1,2,3} has
attracted considerable interest not only in condensed matter physics,
but in quantum field theory as well. This is because at energies below
approximately 3~eV \cite{4} graphene is described by the relativistic
Dirac equation in (2+1) dimensions where the role of the speed of
light $c$ is played by a factor of 300 smaller Fermi velocity $v_F$.
As a result, graphene makes it possible to test the effects of
relativistic quantum field theory, like the Klein paradox \cite{5}
or pair production from vacuum by strong external fields
\cite{6,7,8,9,10,11}, on a laboratory table.

Graphene is unique in that its response functions to the electromagnetic
fluctuations can be expressed via the polarization tensor and
found starting from the first principles of quantum field theory.
There is considerable literature devoted to this subject (see the
list of references in Ref.~\cite{12} where some partial results where
obtained). Finally, the polarization tensor of both pristine and
gapped and doped graphene at any temperature was calculated in Refs.
\cite{13,14}. At nonzero temperature, the results of Ref. \cite{14}
where obtained only at the pure imaginary Matsubara frequencies.
Later on they were analytically continued to the entire complex
frequency plane for the cases of gapped \cite{15} and doped \cite{16}
graphene.

One of the predictions of quantum field theory, which received
widespread attention during the last years, is the Casimir effect
\cite{17}. It is the attractive force acting between two parallel
uncharged material plates in vacuum which is caused by the zero-point
and thermal fluctuations of quantum fields. In the framework of the
Lifshitz theory \cite{18,19}, the Casimir force is expressed through
the reflection coefficients of electromagnetic fluctuations on the
plates. In so doing, both the on- and off-the-mass-shell fluctuations
contribute to the result. For graphene, the exact reflection coefficients
were written in terms of the polarization tensor \cite{13,14}, and the
theoretical predictions of the Lifshitz theory were found to be in a
very good agreement with measurements of the Casimir interaction
\cite{20,21,22,23}.

Quite to the contrary, many measurements of the Casimir force between
metallic and dielectric bodies performed during the last twenty years
were found in disagreement with theoretical predictions of the Lifshitz
theory if the reflection coefficients are expressed via the
universally accepted and well-studied frequency-dependent dielectric
permittivities of the plate materials (see Refs. \cite {24,25,26,27,28}
for a review). The key formal feature of these permittivity functions
is that they possess a simple pole at zero frequency associated
either with the role of conduction electrons in metals or the dc
conductivity of dielectrics (the Drude-like behavior). It was shown
also that an agreement between experiment and theory is restored if
the dielectric permittivity of metallic plates at low frequency is
described by the plasma model possessing a double pole at zero
frequency \cite{24,25,26,27,28}. As to dielectrics, the theoretical
predictions are brought in agreement with the measurement data if
the dc conductivity is omitted in computations, i.e., the regular
at zero frequency dielectric permittivity is used \cite{24,25,26,27,28}.

The surprising thing is that the plasma model does not take into
account the dissipation properties of conduction electrons and it is
not applicable at low frequencies. In a similar way, the dc
conductivity of dielectrics is a really existing effect, and the
theoretical description should not become more precise if we omit it.
Moreover, the Lifshitz theory was found in disagreement with the
third law of thermodynamics (the Nernst heat theorem) for the basic
model of metals with perfect crystal lattices and for all dielectrics
if the Drude-like response functions are used in calculations of the
Casimir force. For the regular or possessing a double pole at zero
frequency response functions, it was proven that the Lifshitz theory
meets the requirements of thermodynamics (see Refs.~\cite{24,25,26,27,28}
for a review).

An important distinction between the response functions of the 3D
materials and graphene is that the former are more or less of the
phenomenological character, whereas the latter are found from the first
principles of quantum field theory. Up to now, precise computations
of the Casimir interaction in graphene systems were based directly
on the polarization tensor, and the closely related to it spatially
nonlocal dielectric permittivities of graphene did not receive due
attention. Keeping in mind, however, that for graphene described by
the polarization tensor the Lifshitz theory is in perfect agreement
with the measurement data \cite{20,21,22,23}, a comparison between the
exact dielectric permittivities of graphene and the phenomenological
permittivities of ordinary materials may be helpful in understanding
the roots of the problems arising for them.

In this paper, we consider the spatially nonlocal dielectric
permittivities of graphene obtained from the polarization tensor in
the areas of both the on-the-mass-shell and off-the-mass-shell waves.
To keep calculations from becoming too involved and to make the
results most transparent, we restrict our attention to the case of a
pristine graphene at zero temperature described by the standard
Dirac model. Both the longitudinal and transverse dielectric
permittivities of graphene are obtained. It is shown that the
longitudinal permittivity is the regular function at zero frequency,
whereas the transverse one possesses at zero frequency a double pole
for any nonzero wave vector.

We compare the forms of dispersion (Kramers-Kronig) relations for the
response functions which are regular at zero frequency or possess
either a simple or a double pole and present several examples from
condensed matter physics and quantum field theory. The dispersion
relations in the forms appropriate for the longitudinal and transverse
permittivities of graphene are proven with account of the spatially
nonlocal effects (previously the Kramers-Kronig relations for the
conductivities of graphene expressed via the polarization tensor
were proven only in the area of propagating waves on the mass shell
where the effects of nonlocality are negligibly small and can be
neglected \cite{29}). The dispersion relations expressing the
permittivities of graphene along the imaginary frequency axis are
also obtained with account of spatial nonlocality. It is shown that
the form of dispersion relations is not affected by the branch points
which are present on the real frequency axis for any nonzero wave
vector. A comparison between computations of the Casimir force for
graphene sheets and metallic plates allows to conclude that the
commonly used dielectric permittivities of metals may be
inapplicable in the area of the off-the-mass-shell electromagnetic
fluctuations.

The paper is organized as follows. In Sec.~II, we compare the
dispersion relations valid for the response functions which are
regular or have a simple or a double pole at zero frequency.
Section III presents the explicit expressions for the polarization
tensor and for the spatially nonlocal longitudinal and transverse
dielectric permittivities of a pristine graphene. The dispersion
relations for the real and imaginary parts of the dielectric
permittivities of graphene with due regard to the off-the-mass-shell
waves are proven in Sec.~IV. Section V is devoted to the dispersion
relations for the permittivities along the imaginary frequency axis.
Section VI contains our conclusions and a discussion of implications
of the obtained results to the Casimir effect. In the Appendices
A and B, several integrals used in Secs.~IV and V are calculated.

\newcommand{\kb}{{k_{\bot}}}
\newcommand{\skb}{{k_{\bot}^2}}
\newcommand{\vk}{{\mbox{\boldmath$k$}}}
\newcommand{\rv}{{\mbox{\boldmath$r$}}}
\newcommand{\ve}{{\varepsilon}}
\newcommand{\rve}{{\rm Re}\,{\varepsilon}}
\newcommand{\ive}{{\rm Im}\,{\varepsilon}}
\newcommand{\ok}{{(\omega,k)}}
\def\Xint#1{\mathchoice
   {\XXint\displaystyle\textstyle{#1}}%
   {\XXint\textstyle\scriptstyle{#1}}%
   {\XXint\scriptstyle\scriptscriptstyle{#1}}%
   {\XXint\scriptscriptstyle\scriptscriptstyle{#1}}%
   \!\int}
\def\XXint#1#2#3{{\setbox0=\hbox{$#1{#2#3}{\int}$}
     \vcenter{\hbox{$#2#3$}}\kern-.5\wd0}}
\def\ddashint{\Xint=}
\def\dashint{\Xint-}
\newcommand{\tR}{{\,-\!\!\!\longrightarrow_{\vphantom{4mm}_{\hspace*{-7mm}\rho\to 0}}{\ }}}

\section{Dispersion relations for the regular and having simple or double poles
response functions}

It is well known that the response functions should be analytic in the upper
half-plane of complex frequencies. This demand is equivalent to the condition
of causality \cite{30}. The response function of a dielectric body to the
electromagnetic field is usually called the electric susceptibility
\begin{equation}
\chi(\omega)=\ve(\omega)-1,
\label{eq1}
\end{equation}
\noindent
where $\ve(\omega)$ is the frequency-dependent dielectric permittivity.
According to the Cauchy theorem, any function $\chi(\omega)$ analytic in the
upper half-plane of complex $\omega$ satisfies the dispersion relations which
are also called the Kramers-Kronig relations. The form of these relations,
however, depends on the properties of $\chi(\omega)$ at the point
$\omega=0$. If  $\chi_{\rm I}(\omega)$ and the corresponding permittivity
 $\ve_{\rm I}(\omega)$ are regular at $\omega=0$, the dispersion relations
take the simplest form \cite{30}
\begin{eqnarray}
&&
\rve_{\rm I}(\omega)-1=\frac{1}{\pi}\dashint_{-\infty}^{\infty}
\frac{\ive_{\rm I}(x)}{x-\omega}dx,
\nonumber \\
&&
\ive_{\rm I}(\omega)=-\frac{1}{\pi}\dashint_{-\infty}^{\infty}
\frac{\rve_{\rm I}(x)}{x-\omega}dx,
\label{eq2}
\end{eqnarray}
\noindent
where the integrals on the right-hand sides should be understood as the principal
values.

The most typical example is the electric susceptibility of an insulator
represented in terms of the set of $K$ oscillators \cite{31}
\begin{equation}
\ve_{\rm I}(\omega)-1=\sum_{j=1}^{K}\frac{g_j}{\omega_j^2-\omega^2-i\gamma_j\omega},
\label{eq3}
\end{equation}
\noindent
where $g_j$ are oscillator strengths, $\omega_j\neq 0$ are the oscillator frequencies,
and $\gamma_j$ are the damping parameters. From Eq.(\ref{eq3}), one obtains the finite
static dielectric permittivity $\ve_{\rm I}(0)<\infty$.

The commonly accepted Drude model describing the conduction electrons in metals
\begin{equation}
\ve_{\rm D}(\omega)-1=-\frac{\omega_p^2}{\omega(\omega+i\gamma)},
\label{eq4}
\end{equation}
\noindent
where $\omega_p$ is the plasma frequency and $\gamma$ is the relaxation parameter,
assumes that the imaginary part of the electric susceptibility has a simple pole
at $\omega=0$. In this case, when deriving the dispersion relations using the
Cauchy theorem,   one should bypass the pole along a semicircle of an infinitely
small radius. As a result, the first equality in Eq.~(\ref{eq2}) remains unchanged,
whereas the second one is replaced with
\begin{equation}
\ive_{\rm D}(\omega)=-\frac{1}{\pi}\dashint_{-\infty}^{\infty}
\frac{\rve_{\rm D}(x)}{x-\omega}dx+\frac{\omega_p^2}{\gamma}\frac{1}{\omega}.
\label{eq5}
\end{equation}
\noindent
The additional term on the right-hand side of Eq.~(\ref{eq5}) shows the
 asymptotic behavior of $\ive_{\rm D}(\omega)$ at $\omega=0$.

The electric susceptibilities having a double pole at zero frequency are not as
widely used as the previous two. Moreover, there are some misleading statements
in the literature concerning these susceptibilities and associated with them
dispersion relations. As mentioned in Sec.~I, the theoretical predictions of
the Lifshitz theory are in agreement with experiments on measuring the Casimir
interaction between metallic test bodies if the low-frequency behavior of the
dielectric permittivity is described by the plasma model
\begin{equation}
\ve_{\rm p}(\omega)-1=-\frac{\omega_p^2}{\omega^2},
\label{eq6}
\end{equation}
\noindent
which has a double pole at zero frequency. This equation is obtained from
Eq.~(\ref{eq4}) by putting $\gamma=0$, i.e., by omitting the dissipation properties
of conduction electrons. These properties are well-studied and play an
important role in numerous physical phenomena but, surprisingly, when included in
the Lifshitz theory, they bring it to a contradiction with the measurement data.

Computations using the Lifshitz theory should take into account both conduction
and bound (core) electrons. This could be made by considering the generalized
Drude- or plasma-like electric susceptibilities where the core electrons are
described by the oscillator term (\ref{eq3}). For example, the generalized
plasma-like susceptibility leading to agreement between the Lifshitz theory and
the measurement data is given by
\begin{equation}
\ve_{\rm gp}(\omega)-1=-\frac{\omega_p^2}{\omega^2}+
\sum_{j=1}^{K}\frac{g_j}{\omega_j^2-\omega^2-i\gamma_j\omega}.
\label{eq7}
\end{equation}

This equation presents an analytic function in the upper half-plane of complex $\omega$.
The real part of this equation has a double pole at $\omega=0$. Therefore, the standard
derivation using the Cauchy theorem with due attention to passing around the point
$\omega=0$ results in the dispersion relations (see Refs.~\cite{25,32} for a detailed
derivation)
\begin{eqnarray}
&&
\rve_{\rm gp}(\omega)-1=\frac{1}{\pi}\dashint_{-\infty}^{\infty}
\frac{\ive_{\rm gp}(x)}{x-\omega}dx-\frac{\omega_p^2}{\omega^2},
\nonumber \\
&&
\ive_{\rm gp}(\omega)=-\frac{1}{\pi}\dashint_{-\infty}^{\infty}
\frac{dx}{x-\omega}\left[\rve_{\rm gp}(x)+
\frac{\omega_p^2}{x^2}\right],
\label{eq8}
\end{eqnarray}

For a simple plasma model (\ref{eq6}), one has $\ive_{\rm p}(\omega)=0$ and,
taking into account that
\begin{equation}
\dashint_{-\infty}^{\infty}\frac{dx}{x-\omega}=0,
\label{eq9}
\end{equation}
\noindent
Eq.~(\ref{eq8}) results in the identities
\begin{eqnarray}
&&
\rve_{\rm p}(\omega)-1=\ve_{\rm p}(\omega)-1=
-\frac{\omega_p^2}{\omega^2},
\nonumber \\
&&
\ive_{\rm p}(\omega)=0.
\label{eq10}
\end{eqnarray}
\noindent
Thus, both functions (\ref{eq6}) and (\ref{eq7}) satisfy the dispersion relations
in the form valid for the electric susceptibilities possessing a double pole at
zero frequency.

In spite of these facts, there are statements in the literature that ``a material
with strictly real $\ve(\omega)$, at all frequencies, is inadmissible. Indeed, such
a material would violate the Kramers-Kronig relations\ldots" \cite{33} and
``a lossless dispersion is incompatible with the Kramers-Kronig relations" \cite{34}.
It was also stated that ``the second order pole cannot exist in any realistic plasma
(even as a meaningful approximation" \cite{33} and ``the second order pole in (6) is
an artifact due to use of a model which is inadmissible at low frequencies" \cite{34}.

The statements of this kind are based on a confusion. It is true that the conduction
electrons in metals possess the dissipation properties. It is not true, however,
that the lossless response functions are incompatible with the Kramers-Kronig
relations. It is the matter of fact of mathematics that any function analytic in the
upper half-plane of complex frequency (including that ones which take real values
along the real frequency axis) satisfy these relations.

As to the response functions possessing the double poles at zero frequency, they are
widely used in the literature. One could mention the Lindhard theory which describes
the screening of electric field by the charge carriers in metals in the random
phase approximation \cite{35}. The transverse dielectric permittivity of a metal
which describes the response to electric field directed perpendicular to the wave
vector obtained in Ref.~\cite{35} has the double pole at $\omega=0$. In doing so
the longitudinal permittivity describing the dielectric response to electric field
parallel to the wave vector remains regular at zero frequency. An example of the
second-order pole in the transverse response function
of an electron gas in the linear
response theory is considered in Ref.~\cite{36}. Mention should be made also of
the dispersion relations for the scattering amplitudes in quantum mechanics and
quantum field theory \cite{37,38}. For a number of processes, the S-matrix and the
scattering amplitudes have the double poles (see, for instance,
Refs.~\cite{39,40,41,42}). These amplitudes satisfy the dispersion relations with
appropriate subtractions.

The phenomenological spatially nonlocal transverse permittivities, which possess the
double pole at zero frequency for a nonzero wave vector and coincide with the Drude
model (\ref{eq4}) for a vanishing wave vector, were suggested in Refs.~\cite{43,44}.
It was shown that these permittivities satisfy the Kramers-Kronig relations and bring
the Lifshitz theory in agreement with the measurement data of all experiments on
measuring the Casimir force \cite{43,44,45}, as well as with the requirements of
thermodynamics \cite{46}. Note that Ref.~\cite{33}
also underlines that at low frequencies one should take into account the effects of
spatial dispersion. However, the specific spatially nonlocal dielectric function
derived from the kinetic theory considered in Ref.~\cite{33} does not bring the Lifshitz
theory in agreement with the measurement data.

That is why in the next sections we analyze the analytic properties of response
functions and the form of dispersion relations for graphene where all the results
are obtained on the solid foundation of quantum field theory do not using any
phenomenology.

\section{The polarization tensor and the spatially nonlocal
dielectric permittivities of graphene described by the Dirac model}

The polarization tensor of graphene $\Pi_{\mu\nu}\ok$, where $\mu,\,\nu=0,\,1,\,2$,
$\omega$ is the frequency, and $k$ is the magnitude of the two-dimensional wave
vector, was calculated in Refs.~\cite{13,14} at the pure imaginary Matsubara
frequencies in the one-loop approximation and analytically continued to the
entire complex frequency plane in Ref.~\cite{15}. This tensor can be
expressed via the two independent quantities, for instance, $\Pi_{00}\ok$ and
$\Pi^{\rm tr}\ok=\Pi_\mu^{\,\mu}\ok$. For our purposes it is more  convenient
to use the combination
\begin{equation}
\Pi\ok=k^2\Pi^{\rm tr}\ok+\frac{1}{c^2}(\omega^2-c^2k^2)\Pi_{00}\ok
\label{eq11}
\end{equation}
\noindent
instead of $\Pi^{\rm tr}\ok$. Below we consider the simplest case of a pristine
(undoped and ungapped) graphene at zero temperature.

The specific expressions for the polarization tensor depend on the frequency region
under consideration. Thus, for $-v_Fk<\omega<v_Fk$ (we recall that $v_F\approx c/300$
is the Fermi velocity for graphene) it holds \cite{13,14,15}
\begin{eqnarray}
&&
\Pi_{00}\ok=\frac{\pi\alpha\hbar k^2c}{\sqrt{v_F^2k^2-\omega^2}},
\nonumber\\
&&
\Pi\ok=\pi\alpha\hbar\,\frac{k^2}{c}{\sqrt{v_F^2k^2-\omega^2}},
\label{eq12}
\end{eqnarray}
\noindent
where $\alpha=e^2/(\hbar c)\approx 1/137$ is the fine structure constant.

In the frequency regions $\omega>v_Fk$ and $\omega<-v_Fk$ one obtains \cite{13,14,15}
\begin{eqnarray}
&&
\Pi_{00}\ok=\pm i\frac{\pi\alpha\hbar k^2c}{\sqrt{\omega^2-v_F^2k^2}},
\nonumber\\
&&
\Pi\ok=\mp i\pi\alpha\hbar\,\frac{k^2}{c}{\sqrt{\omega^2-v_F^2k^2}},
\label{eq13}
\end{eqnarray}
\noindent
where the upper and lower signs stand for the positive and negative $\omega$,
respectively.

The polarization tensor is directly connected with the spatially nonlocal electric
susceptibilities, density-density correlation functions, and dielectric permittivities
of graphene. Thus, the longitudinal and transverse electric susceptibilities
and permittivities of graphene are
expressed as \cite{47,48,49}
\begin{eqnarray}
&&
\ve^{\rm L}\ok-1=\frac{1}{2\hbar k}\Pi_{00}\ok,
\nonumber \\
&&
\ve^{\rm Tr}\ok-1=-\frac{c^2}{2\hbar k\omega^2}\Pi\ok.
\label{eq14}
\end{eqnarray}

Substituting the first equalities of Eqs.~(\ref{eq12}) and (\ref{eq13}) in the first
line of Eq.~({\ref{eq14}), we find expressions for the longitudinal electric
susceptibility and dielectric permittivity of graphene in different frequency regions
\begin{equation}
\ve^{\rm L}\ok-1=\left\{
\begin{array}{ll}
\frac{\pi\alpha kc}{2\sqrt{v_F^2k^2-\omega^2}}, & |\omega|<v_Fk, \\
\pm i\frac{\pi\alpha kc}{2\sqrt{\omega^2-v_F^2k^2}}, & |\omega|>v_Fk.
\end{array}
\right.
\label{eq15}
\end{equation}
\noindent
In the second line of this equation, the sign plus stands for the positive $\omega$
($\omega>v_Fk$) and the sign minus stands for the negative $\omega$
($\omega<-v_Fk$).

As is seen from Eq.~(\ref{eq15}), the longitudinal dielectric permittivity of graphene
is regular at zero frequency for any wave vector. Thus, in this respect it is somewhat
similar to the permittivity of an insulator.

The transverse electric susceptibility and dielectric permittivity of graphene deserve
a closer examination.
Substituting the second equalities of Eqs.~(\ref{eq12}) and (\ref{eq13}) in the second
line of Eq.~({\ref{eq14}), we obtain expressions for the transverse electric
susceptibility and dielectric permittivity of graphene
\begin{equation}
\ve^{\rm Tr}\ok-1=\left\{
\begin{array}{ll}
-\frac{\pi\alpha kc}{2\omega^2}\sqrt{v_F^2k^2-\omega^2}, & |\omega|<v_Fk, \\
\pm i\frac{\pi\alpha kc}{2\omega^2}\sqrt{\omega^2-v_F^2k^2}, & |\omega|>v_Fk,
\end{array}
\right.
\label{eq16}
\end{equation}
\noindent
where again the upper sign in the second line of this equation stands for the positive $\omega$
 and the lower sign stands for the negative $\omega$.

As is seen from Eq.~(\ref{eq16}), the real part of the transverse electric susceptibility and
dielectric permittivity of graphene for any nonzero  wave vector $k$ possesses the double pole
at zero frequency. In this case, the presence of the double pole in the response function
is a direct consequence of the quantum field theoretical formalism without resorting to
any phenomenological approach.

The obtained dielectric permittivities of graphene are the analytic functions in the upper
half-plane of complex frequency. The real and imaginary parts of these permittivities
are the even and odd functions under the change of the  sign of frequency, respectively,
with unchanged $k$
\begin{eqnarray}
&&
\rve^{\rm L,Tr}\ok=\rve^{\rm L,Tr}(-\omega,k),
\nonumber\\
&&
\ive^{\rm L,Tr}\ok=-\ive^{\rm L,Tr}(-\omega,k),
\label{eq17}
\end{eqnarray}
\noindent
as it should be for the nonlocal response functions \cite{30}.
For the positive $\omega$, both $\ive^{\rm L}\ok$ and $\ive^{\rm Tr}\ok$ are positive.
The branch points which are present in both $\ve^{\rm L}\ok$ and $\ve^{\rm Tr}\ok$
at $\omega=\pm v_Fk$ for any nonzero $k$ are considered in Sec.~V.
In the next section, we elucidate the form of dispersion relations satisfied by the
real and imaginary parts of the response functions of graphene.

\section{The dispersion relations for graphene with regard to the
off-the-mass-shell waves}

The response functions of graphene
 (\ref{eq15}) and (\ref{eq16}) are written for both
the on- and off-the-mass-shell waves. In Ref.~\cite{29} the Kramers-Kronig relations
for the conductivity of graphene were obtained only for the propagating waves on
the mass shell which satisfy the condition $\omega>kc$. In this case
\begin{eqnarray}
&&
\sqrt{\omega^2-v_F^2k^2}=\omega\sqrt{1-\left(\frac{v_Fk}{\omega}\right)^2}
\nonumber \\
&&~~~~~~~~~~~~
=
\omega\sqrt{1-\left(\frac{v_F}{c}\right)^2\left(\frac{kc}{\omega}\right)^2}
\approx\omega
\label{eq18}
\end{eqnarray}
\noindent
and one can neglect by the effects of spatial nonlocality. Below we consider the
response functions to the on- and off-the-mass-shell waves on equal terms.

We start with the most interesting case of the transverse dielectric permittivity,
$\ve^{\rm Tr}\ok$, which possesses a double pole at zero frequency. In this case,
according to Eq.~(\ref{eq16}), for the real part of  $\ve^{\rm Tr}\ok$ one has
\begin{equation}
\rve^{\rm Tr}\ok=\left\{
\begin{array}{ll}
1-\frac{\pi\alpha kc}{2\omega^2}\sqrt{v_F^2k^2-\omega^2}, & |\omega|<v_Fk, \\
1, & |\omega|>v_Fk,
\end{array}
\right.
\label{eq19}
\end{equation}
\noindent
The imaginary part of  $\ve^{\rm Tr}\ok$ takes the form
\begin{equation}
\ive^{\rm Tr}\ok=\left\{
\begin{array}{ll}
0, & |\omega|<v_Fk, \\
 \frac{\pi\alpha kc}{2\omega^2}\sqrt{\omega^2-v_F^2k^2}, & \omega>v_Fk, \\
-\frac{\pi\alpha kc}{2\omega^2}\sqrt{\omega^2-v_F^2k^2}, & \omega<-v_Fk.
\end{array}
\right.
\label{eq20}
\end{equation}

At first we consider the dispersion relation expressing the real part of $\ve^{\rm Tr}\ok$
via its imaginary part. Using a similarity with $\ve_{\rm gp}(\omega)$ in the dispersion
relation (\ref{eq8}), which is  valid for the permittivity possessing the double pole
at $\omega=0$, we consider the function
\begin{equation}
F^{\rm Tr}\ok-1=\frac{1}{\pi}\dashint_{-\infty}^{\infty}\!
\frac{\ive^{\rm Tr}(x,k)}{x-\omega}\,dx-\frac{\pi\alpha k^2cv_F}{2\omega^2}.
\label{eq21}
\end{equation}
\noindent
Similar to Eq.~(\ref{eq8}), the last term on the right-hand side of Eq.~(\ref{eq21})
presents the asymptotic behavior of $\rve^{\rm Tr}\ok$ from Eq.~(\ref{eq19}) in the
limiting case $\omega\to 0$.

Now we substitute Eq.~(\ref{eq20}) in Eq.~(\ref{eq21}) and obtain
\begin{eqnarray}
&&
F^{\rm Tr}\ok-1=\frac{\alpha kc}{2}\left[-\dashint_{-\infty}^{-b}
\frac{dx\sqrt{x^2-b^2}}{x^2(x-\omega)}\right.
\nonumber \\
&&~~~~
\left.+\dashint_{b}^{\infty}
\frac{dx\sqrt{x^2-b^2}}{x^2(x-\omega)}\right]
-\frac{\pi\alpha k^2cv_F}{2\omega^2},
\label{eq22}
\end{eqnarray}
\noindent
where the notation $b\equiv v_Fk$ is introduced.

By changing the sign of the integration variable in the first integral on the
right-hand side of Eq.~(\ref{eq22}), after identical transformations, we find
\begin{equation}
F^{\rm Tr}\ok-1={\alpha kc}\left[\dashint_{b}^{\infty}\!
\frac{dx\sqrt{x^2-b^2}}{x(x^2-\omega^2)}
-\frac{\pi b}{2\omega^2}\right].
\label{eq23}
\end{equation}
\noindent
Here, we introduce the integration variable $y=x^2-b^2$ and rewrite
Eq.~(\ref{eq23}) as
\begin{equation}
F^{\rm Tr}\ok-1=\frac{\alpha kc}{2}\left[\dashint_{0}^{\infty}\!
\frac{\sqrt{y}\,dy}{(y+b^2)(y+b^2-\omega^2)}
-\frac{\pi b}{\omega^2}\right].
\label{eq24}
\end{equation}

Under the condition $|\omega|<b$ this integral is easily calculated using the result
3.223(1) in Ref.~\cite{50} leading to
\begin{equation}
F^{\rm Tr}\ok-1=-\frac{\pi\alpha kc}{2\omega^2}\sqrt{b^2-\omega^2},
\label{eq25}
\end{equation}
\noindent
which is in agreement with the first line of Eq.~(\ref{eq19}) giving the real part
of $\ve^{\rm Tr}$ for $|\omega|<b$.

If the opposite condition $|\omega|>b$ is satisfied, one can use the integral
3.223(2) in Ref.~\cite{50} with the result
\begin{equation}
F^{\rm Tr}\ok-1=0
\label{eq26}
\end{equation}
\noindent
in agreement  with the second line of Eq.~(\ref{eq19}).

Thus, the dispersion relation
\begin{equation}
\rve^{\rm Tr}\ok-1=\frac{1}{\pi}\dashint_{-\infty}^{\infty}\!
\frac{\ive^{\rm Tr}(x,k)}{x-\omega}\,dx-\frac{\pi\alpha k^2cv_F}{2\omega^2}
\label{eq27}
\end{equation}
\noindent
is finally proven.

We are coming now to the inverse dispersion relation expressing the imaginary part
of $\ve^{\rm Tr}\ok$ via its real part. Taking again into account the similarity
with the dielectric function $\ve_{\rm gp}$ in Eq.~(\ref{eq8}), we consider the
quantity
\begin{equation}
G^{\rm Tr}\ok=-\frac{1}{\pi}\dashint_{\!-\infty}^{\infty}\!
\frac{dx}{x-\omega}\left[
\rve^{\rm Tr}(x,k)+\frac{\pi\alpha k^2cv_F}{2x^2}\right].
\label{eq28}
\end{equation}

Substituting here Eq.~(\ref{eq19}) with account of Eq.~(\ref{eq9}), one obtains
\begin{equation}
G^{\rm Tr}\ok=\frac{\alpha kc}{2}\left[\dashint_{\!-b}^{b}\!
\frac{dx\sqrt{b^2-x^2}}{x^2(x-\omega)}
-b \dashint_{\!-\infty}^{\infty}\!\frac{dx}{x^2(x-\omega)}\right].
\label{eq29}
\end{equation}

Now we use the identity
\begin{equation}
\frac{1}{x^2(x-\omega)}=\frac{1}{\omega^2(x-\omega)}-\frac{1}{\omega^2x}-
\frac{1}{\omega x^2}
\label{eq30}
\end{equation}
\noindent
in the second integral on the right-hand side of Eq.~(\ref{eq29}). Taking into
consideration Eq.~(\ref{eq9}) in its immediate form and with $\omega=0$, one finds
\begin{eqnarray}
&&
\dashint_{\!-\infty}^{\infty}\!\frac{dx}{x^2(x-\omega)}=
-\frac{1}{\omega}\dashint_{\!-\infty}^{\infty}\!\frac{dx}{x^2}
\nonumber \\
&&~~~~~~~~~~~
=-\frac{1}{\omega}\left(\dashint_{\!-b}^{b}\!\frac{dx}{x^2}+
\frac{2}{b}\right).
\label{eq31}
\end{eqnarray}

Substituting Eq.~(\ref{eq30}) in the first integral on the right-hand side of
Eq.~(\ref{eq29}) and using Eq.~(\ref{eq31}), we arrive at
\begin{eqnarray}
&&
G^{\rm Tr}\ok=\frac{\alpha kc}{2}\left(\frac{1}{\omega^2}\dashint_{\!-b}^{b}\!
\frac{dx\sqrt{b^2-x^2}}{x-\omega}\right.
\nonumber \\
&&~~~~~~\left.
+\frac{1}{\omega}\dashint_{\!-b}^{b}\!\!dx\frac{b-\sqrt{b^2-x^2}}{x^2}+
\frac{2}{\omega}\right),
\label{eq32}
\end{eqnarray}
\noindent
where it was taken into account that
\begin{equation}
\dashint_{\!-b}^{b}\!\!dx\frac{\sqrt{b^2-x^2}}{x}=0
\label{eq33}
\end{equation}
\noindent
as an integral of the odd function over the symmetric interval.

The two integrals on the right-hand side of Eq.~(\ref{eq32}) are calculated in the
Appendix A. Substituting Eqs.~(\ref{A6}) and (\ref{A9}) in Eq.~(\ref{eq32}), we
finally obtain
\begin{equation}
G^{\rm Tr}\ok=\frac{\alpha kc}{2}\left\{
\begin{array}{ll}
0, & |\omega|<b, \\
\frac{\pi\sqrt{\omega^2-b^2}}{\omega^2}, & \omega>b, \\
-\frac{\pi\sqrt{\omega^2-b^2}}{\omega^2}, & \omega<-b.
\end{array}
\right.
\label{eq34}
\end{equation}
\noindent
These results are in agreement with the imaginary part of the transverse dielectric
permittivity of graphene $\ive^{\rm Tr}\ok$ in Eq.~(\ref{eq20}). Thus, the inverse
dispersion relation takes the form
\begin{equation}
\ive^{\rm Tr}\ok=-\frac{1}{\pi}\dashint_{\!-\infty}^{\infty}\!
\frac{dx}{x-\omega}\left[
\rve^{\rm Tr}(x,k)+\frac{\pi\alpha k^2cv_F}{2x^2}\right].
\label{eq35}
\end{equation}

We are coming now to the dispersion relations for the longitudinal dielectric
permittivity of graphene $\ve^{\rm L}\ok$. According to Eq.~(\ref{eq15}), for the
real part of this permittivity one has
\begin{equation}
\rve^{\rm L}\ok=\left\{
\begin{array}{ll}
1+\frac{\pi\alpha kc}{2\sqrt{v_F^2k^2-\omega^2}}, & |\omega|<v_Fk, \\
1, & |\omega|>v_Fk,
\end{array}
\right.
\label{eq36}
\end{equation}
\noindent
whereas for its imaginary part one obtains
\begin{equation}
\ive^{\rm L}\ok=\left\{
\begin{array}{ll}
0, & |\omega|<v_Fk, \\
 \frac{\pi\alpha kc}{2\sqrt{\omega^2-v_F^2k^2}}, & \omega>v_Fk, \\
-\frac{\pi\alpha kc}{2\sqrt{\omega^2-v_F^2k^2}}, & \omega<-v_Fk.
\end{array}
\right.
\label{eq37}
\end{equation}

This permittivity is regular at zero frequency. Because of this, using the similarity
with $\ve_{\rm I}(\omega)$ in Eq.~(\ref{eq2}), we consider the function
\begin{equation}
F^{\rm L}\ok-1=\frac{1}{\pi}\dashint_{-\infty}^{\infty}\!
\frac{\ive^{\rm L}(x,k)}{x-\omega}dx.
\label{eq38}
\end{equation}

Substituting Eq.~(\ref{eq37}) in Eq.~(\ref{eq38}), we find
\begin{eqnarray}
&&
F^{\rm L}\ok-1=\frac{\alpha kc}{2}\left[-\dashint_{-\infty}^{-b}\!
\frac{dx}{(x-\omega)\sqrt{x^2-b^2}}\right.
\nonumber \\
&&~~~~~~~~~\left.
+\dashint_{b}^{\infty}\!
\frac{dx}{(x-\omega)\sqrt{x^2-b^2}}\right].
\label{eq39}
\end{eqnarray}
\noindent
By changing the sign of the integration variable in the first integral of this equation,
after identical transformations we bring it to the form
\begin{equation}
F^{\rm L}\ok-1={\alpha kc}\dashint_{b}^{\infty}\!
\frac{xdx}{(x^2-\omega^2)\sqrt{x^2-b^2}}.
\label{eq40}
\end{equation}

Now we introduce the integration variable $u=x^2-b^2$ and obtain
\begin{equation}
F^{\rm L}\ok-1=\frac{\alpha kc}{2}\dashint_{0}^{\infty}\!
\frac{du}{(u+b^2-\omega^2)\sqrt{u}}.
\label{eq41}
\end{equation}
\noindent
Evaluating the last integral with the help of 3.222(2) in Ref.~\cite{50},
we finally find
\begin{equation}
F^{\rm L}\ok-1=\left\{
\begin{array}{ll}
\frac{\pi\alpha kc}{2\sqrt{b^2-\omega^2}}, & |\omega|<b, \\
0, & |\omega|>b.
\end{array}
\right.
\label{eq42}
\end{equation}

These results agree with the real part of the longitudinal permittivity of graphene
in Eq.~(\ref{eq36}). For this reason, the first dispersion relation takes the form
\begin{equation}
\rve^{\rm L}\ok-1=\frac{1}{\pi}\dashint_{-\infty}^{\infty}\!
\frac{\ive^{\rm L}(x,k)}{x-\omega}dx.
\label{eq43}
\end{equation}

To prove the validity of the inverse dispersion relation for $\ve^{\rm L}\ok$,
we consider the quantity
\begin{equation}
G^{\rm L}\ok=-\frac{1}{\pi}\dashint_{-\infty}^{\infty}\!
\frac{\rve^{\rm L}(x,k)}{x-\omega}dx.
\label{eq44}
\end{equation}

Substituting here the real part of $\ve^{\rm L}\ok$ from Eq.~(\ref{eq36}),
one obtains
\begin{eqnarray}
&&
G^{\rm L}\ok=-\frac{1}{\pi}\dashint_{-b}^{b}\!
\frac{dx}{x-\omega}\left(1+\frac{\pi\alpha kc}{2\sqrt{b^2-x^2}}\right)
\nonumber \\
&&~~~~~~~~~~
-\frac{1}{\pi}\left(\dashint_{b}^{\infty}\!
\frac{dx}{x-\omega}+\dashint_{-\infty}^{-b}\!
\frac{dx}{x-\omega}\right).
\label{eq45}
\end{eqnarray}
\noindent
Combining the last two integrals with the first contribution to the first integral
and taking into account Eq.~(\ref{eq9}), we simplify  Eq.~(\ref{eq45}) to
\begin{eqnarray}
&&
\hspace*{-5mm}
G^{\rm L}\ok=-\frac{\alpha kc}{2}\dashint_{-b}^{b}\!
\frac{dx}{(x-\omega)\sqrt{b^2-x^2}}
\label{eq46} \\
&&
\hspace*{-5mm}
=-\frac{\alpha kc}{2}
\left[\dashint_{-b}^{0}\!\frac{dx}{(x-\omega)\sqrt{b^2-x^2}}
+\dashint_{0}^{b}\!\frac{dx}{(x-\omega)\sqrt{b^2-x^2}}
\right].
\nonumber
\end{eqnarray}
\noindent
By changing the sign of the integration variable in the first integral,
after identical transformations, we rewrite Eq.~(\ref{eq46}) as
\begin{equation}
G^{\rm L}\ok=\alpha kc\omega\dashint_{0}^{b}\!
\frac{dx}{(\omega^2-x^2)\sqrt{b^2-x^2}}.
\label{eq47}
\end{equation}

This integral is calculated using the result 1.2.50(10) in Ref.~\cite{51}
\begin{equation}
G^{\rm L}\ok=\frac{\alpha kc}{2}\left\{
\begin{array}{ll}
0, &|\omega|<b, \\
\frac{\pi}{\sqrt{\omega^2-b^2}}, & \omega>b, \\
-\frac{\pi}{\sqrt{\omega^2-b^2}}, & \omega<-b.
\end{array}
\right.
\label{eq48}
\end{equation}

By comparing Eq.~(\ref{eq48}) with Eq.~(\ref{eq37}), one arrives to the inverse
dispersion relation for the longitudinal dielectric permittivity of graphene
\begin{equation}
\ive^{\rm L}\ok=-\frac{1}{\pi}\dashint_{-\infty}^{\infty}\!
\frac{\rve^{\rm L}(x,k)}{x-\omega}dx.
\label{eq49}
\end{equation}

Hence the spatially nonlocal dielectric permittivities of graphene $\ve^{\rm Tr}\ok$
and $\ve^{\rm L}\ok$ satisfy the dispersion relations given by Eqs.~(\ref{eq27}) and
(\ref{eq35}) and by Eqs.~(\ref{eq43}) and (\ref{eq49}), respectively. These
dispersion relations have the same form as in the case of spatially local
permittivities having a similar pole structure at zero frequency but depend on the
wave vector magnitude as a parameter. This is in accordance with the standard
approach of classical electrodynamics of continuous media \cite{30}.

The additional terms in the dispersion relations for graphene (\ref{eq27}) and
(\ref{eq35}) originate from the double pole at zero frequency, which is present
in the real part of the transverse dielectric permittivity $\ve^{\rm Tr}\ok$.
This pole is in some formal analogy to that in the generalized plasma-like
permittivity (\ref{eq7}). As a result, the dispersion relations (\ref{eq8}),
on the one hand, and
(\ref{eq27}) and (\ref{eq35}), on the other hand,
have a similar form. It should be remembered,
however, that the pole structure of the response functions of graphene is derived
from the first principles of quantum field theory, whereas the term
$-\omega_p^2/\omega^2$ in the permittivity (\ref{eq7}) was introduced in a
phenomenological manner by omitting the dissipation properties of conduction
electrons in order to bring the Lifshitz theory in agreement with the measurement
data (see Sec.~VI for a possible role of these results for resolving
problems in theoretical description of the Casimir force between metallic plates).
Note also that the spatially nonlocal permittivities of graphene are nonanalytic
at the branch points $\omega=\pm v_Fk$ on the real frequency axis. As is seen from
the above and from the next section, this, however, does not affect the form
of dispersion relations.

\section{The dispersion relations for the response functions of graphene along the
imaginary frequency axis}

Here, we derive the dispersion relations representing the dielectric permittivities
of graphene along the imaginary frequency axis. For this purpose, let us consider
the integral
\begin{equation}
\int_C\frac{\omega[\ve\ok-1]}{\omega^2+\xi^2}d\omega,
\label{eq50}
\end{equation}
\noindent
where $\ve\ok$ is either the transverse or the longitudinal permittivity of graphene.
The contour $C$ in the plane of complex $\omega$ consists of a semicircle $C_R$ of
the infinitely large radius $R$, three semicircles $C_{\rho}^{\,l}$, $C_{\rho}^{\,r}$
and $C_{\rho}$ of the infinitely small radii $\rho$ around the branch points
$\omega=\mp v_Fk$ and $\omega=0$, and the real frequency axis (see Fig.~1).
The dashed line in Fig.~1 shows the lower edge of the branch cut between the points
$\omega=\mp v_Fk$.
\begin{figure}[!t]
\vspace*{-11.cm}
\centerline{\hspace*{1.5cm}
\includegraphics[width=6.0in]{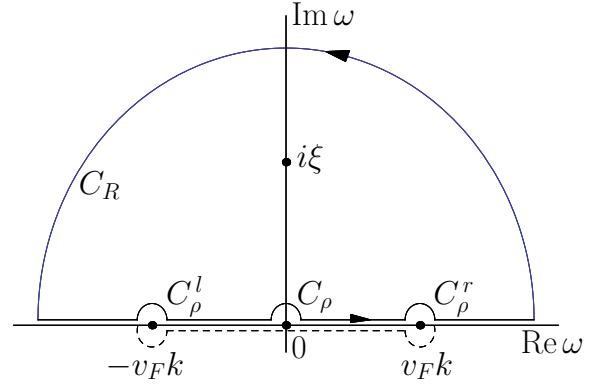}}
\vspace*{-5.cm}
\caption{\label{fg1}The contour of integration $C$ in the upper half-plane of
complex frequency consisting of the real frequency axis, the
semicircle $C_R$ of an infinitely large radius $R$, the semicircles
$C_{\rho}^{\,l}$ and $C_{\rho}^{\,r}$ of the infinitely small radii $\rho$
around the branch points at $\omega = \mp v_{F}k$, and the semicircle
$C_{\rho}$ of an infinite small radius $\rho$ around the double pole
at $\omega=0$. The lower edge of the branch cut is shown by the dashed
line.}
\end{figure}


Inside of the contour $C$ the function under the integral in Eq.~(\ref{eq50}) possesses
the single simple pole at the point $\omega=i\xi$ of the imaginary frequency axis.
Because of this, the integral (\ref{eq50}) is calculated by using the Cauchy's
residue theorem
\begin{eqnarray}
&&
\int_C\frac{\omega[\ve\ok-1]}{\omega^2+\xi^2}d\omega=
2\pi i{\rm Res}{\hspace*{-6mm}{\vphantom{\Bigl(}}_{\omega=i\xi}}
\frac{\omega[\ve\ok-1]}{\omega^2+\xi^2}
\nonumber \\
&&~~~~~~~~~~~~~~
=\pi i[\ve(i\xi,k)-1].
\label{eq51}
\end{eqnarray}

The integral on the left-hand side of Eq.~(\ref{eq51}) takes different values
for $\ve\ok=\ve^{\rm Tr}\ok$ and  $\ve\ok=\ve^{\rm L}\ok$. We begin with the first
option and consider
\begin{equation}
H^{\rm Tr}(\xi,k)=\int_C\frac{\omega[\ve^{\rm Tr}\ok-1]}{\omega^2+\xi^2}d\omega.
\label{eq52}
\end{equation}

The quantity $H^{\rm Tr}(\xi,k)$ can be presented as the sum of the integrals along
the real frequency axis from $-\infty$ to $+\infty$ and along the contours
$C_{\rho}^{\,l}$, $C_{\rho}$, $C_{\rho}^{\,r}$, and $C_R$. In so doing, the contour
$C_R$ should be bypassed in the positive direction (i.e., counter clockwise)
whereas the semicircles $C_{\rho}^{\,l}$, $C_{\rho}$, and $C_{\rho}^{\,r}$ are
bypassed in the negative direction (i.e., clockwise).

It is easily seen that the integral along the contour $C_R$ vanishes. Let us
calculate
\begin{equation}
H_{C_{\rho}}^{\rm Tr}(\xi,k)=\int_{C_{\rho}}
\frac{\omega[\ve^{\rm Tr}\ok-1]}{\omega^2+\xi^2}d\omega,
\label{eq53}
\end{equation}
\noindent
where $\ve^{\rm Tr}\ok-1$ is explicitly defined by the first line in Eq.~(\ref{eq16}).
Substituting this explicit expression in Eq.~({\ref{eq53}), one obtains
\begin{equation}
H_{C_{\rho}}^{\rm Tr}(\xi,k)=-\frac{\pi\alpha kc}{2}\int_{C_{\rho}}
\frac{\sqrt{b^2-\omega^2}\,d\omega}{\omega(\omega^2+\xi^2)}.
\label{eq54}
\end{equation}

The semicircle $C_{\rho}$ can be presented in the form $\omega=\rho e^{i\varphi}$ where
$\varphi$ varies from $\pi$ to 0. Then Eq. ({\ref{eq54}) is rewritten as
\begin{eqnarray}
&&
H_{C_{\rho}}^{\rm Tr}(\xi,k)=-i\frac{\pi\alpha kc}{2}\int_{\pi}^{0}
\frac{\sqrt{b^2-\rho^2e^{2i\varphi}}}{\rho^2e^{2i\varphi}+\xi^2}d\varphi
\nonumber \\
&&~~~
\tR  \> i\frac{\pi\alpha kc}{2}\int_{0}^{\pi}
\frac{b}{\xi^2}d\varphi=i\frac{\pi^2\alpha kcb}{2\xi^2}.
\label{eq55}
\end{eqnarray}

In the Appendix B it is proven that
\begin{equation}
\int_{C_{\rho}^{\,l}}\!\!
\frac{\omega[\ve^{\rm Tr}\ok-1]}{\omega^2+\xi^2}d\omega=
\int_{C_{\rho}^{\,r}}\!\!
\frac{\omega[\ve^{\rm Tr}\ok-1]}{\omega^2+\xi^2}d\omega=0,
\label{eq56}
\end{equation}
\noindent
i.e., the branch points $\omega=\pm v_Fk$ do not contribute to the result.
Substituting Eqs.~(\ref{eq55}) and (\ref{eq56}) in Eq.~(\ref{eq51}) written for
$\ve\ok=\ve^{\rm Tr}\ok$, we find
\begin{equation}
\ve^{\rm Tr}(i\xi,k)-1=-\frac{i}{\pi}\dashint_{-\infty}^{\infty}
\frac{\omega[\ve^{\rm Tr}\ok-1]}{\omega^2+\xi^2}d\omega+
\frac{\pi\alpha k^2cv_F}{2\xi^2}.
\label{eq57}
\end{equation}

Taking into account that the following integrals of the odd functions of
$\omega$ vanish
\begin{equation}
\dashint_{0}^{\infty}
\frac{\omega\,d\omega}{\omega^2+\xi^2}=
\dashint_{0}^{\infty}
\frac{\omega\rve^{\rm Tr}\ok}{\omega^2+\xi^2}d\omega=0,
\label{eq58}
\end{equation}
\noindent
we rewrite Eq.~(\ref{eq57}) in the form
\begin{equation}
\ve^{\rm Tr}(i\xi,k)-1=\frac{2}{\pi}\dashint_{0}^{\infty}
\frac{\omega\ive^{\rm Tr}\ok}{\omega^2+\xi^2}d\omega+
\frac{\pi\alpha k^2cv_F}{2\xi^2},
\label{eq59}
\end{equation}
\noindent
which is the final form of the dispersion relation expressing $\ve^{\rm Tr}(i\xi,k)$
via $\ive^{\rm Tr}\ok$. The last term on the right-hand side of the Eq.~(\ref{eq59})
originates from the double pole of $\ve^{\rm Tr}\ok$ at zero frequency.

We are coming now to the longitudinal dielectric permittivity of graphene
$\ve^{\rm L}\ok$ and consider
\begin{equation}
H^{\rm L}(\xi,k)=\int_C\!\!
\frac{\omega[\ve^{\rm L}\ok-1]}{\omega^2+\xi^2}d\omega,
\label{eq60}
\end{equation}
\noindent
where the contour $C$ is shown in Fig.~1.
The integral (\ref{eq60}) is again presented as the sum of the integrals along
the real frequency axis  and along the contours
$C_{\rho}^{\,l}$, $C_{\rho}$, $C_{\rho}^{\,r}$, and $C_R$ with the vanishing integral
$H_{C_R}^{\rm L}$ along the latter in the limiting case $R\to\infty$.

For the dielectric permittivity  $\ve^{\rm L}\ok$ given by the first line of
Eq.~(\ref{eq15}), the point $\omega=0$ is regular. Because of this
\begin{equation}
H_{\rho}^{\rm L}(\xi,k)=\int_{C_{\rho}}\!\!
\frac{\omega[\ve^{\rm L}\ok-1]}{\omega^2+\xi^2}d\omega
\tR 0.
\label{eq61}
\end{equation}
\noindent
The explicit calculation using Eq.~(\ref{eq15}) confirms this conclusion.

According to Eq.~(\ref{eq15}),  at the branch points $\omega=\pm v_Fk$ the permittivity
$\ve^{\rm L}\ok$ diverges by taking the real and complex values depending on
whether the approach to a singular point along the real frequency axis occurs from
the smaller or larger in magnitude values of frequency. In spite of this fact,
as shown in the Appendix B,
\begin{equation}
\int_{C_{\rho}^{\,l}}\!\!
\frac{\omega[\ve^{\rm L}\ok-1]}{\omega^2+\xi^2}d\omega=
\int_{C_{\rho}^{\,r}}\!\!
\frac{\omega[\ve^{\rm L}\ok-1]}{\omega^2+\xi^2}d\omega=0,
\label{eq62}
\end{equation}
\noindent
i.e., the branch points again do not contribute to the result.

Thus, using Eq.~(\ref{eq51}) written in this case for  $\ve^{\rm L}\ok$,
one obtains
\begin{equation}
\ve^{\rm L}(i\xi,k)-1=-\frac{i}{\pi}\dashint_{-\infty}^{\infty}\!\!
\frac{\omega[\ve^{\rm L}\ok-1]}{\omega^2+\xi^2}d\omega.
\label{eq63}
\end{equation}

With the help of Eq.~(\ref{eq58}), where  $\rve^{\rm Tr}\ok$ is replaced with
$\rve^{\rm L}\ok$, this equation can be rewritten in the form
\begin{equation}
\ve^{\rm L}(i\xi,k)-1=\frac{2}{\pi}\dashint_{0}^{\infty}\!\!
\frac{\omega\ive^{\rm L}\ok}{\omega^2+\xi^2}d\omega,
\label{eq64}
\end{equation}
\noindent
which is the standard form of the dispersion relation valid for the response
functions which are regular at zero frequency. Thus, the presence of the branch
points and respective cut shown in Fig.~1 in the case of graphene makes no impact
on the form of dispersion relations. Because of this, the statement that the
dielectric permittivity has no singular points on the real frequency axis with the
possible exception of only the coordinate origin \cite{30} is, broadly speaking,
inapplicable in the presence of spatial dispersion.

\section{Conclusions and discussion of implications to the Casimir effect}

In the foregoing, we have investigated the spatially nonlocal
longitudinal and transverse dielectric permittivities of graphene
expressed via the polarization tensor based on the first principles
of quantum field theory. It was shown that at zero frequency the
longitudinal permittivity is the regular function whereas the
transverse one possesses a double pole for any nonzero wave vector.
The obtained expressions are valid for any relationship between the
frequency and the wave vector and, thus, describe the electromagnetic
response of graphene to both the on-the-mass-shell and
off-the-mass-shell fields.

According to our results, both the transverse and longitudinal
permittivities of graphene are the analytic functions in the upper
half-plane of complex frequency and satisfy the dispersion
(Kramers-Kronig) relations for their real and imaginary parts for
any value of the wave vector. In doing so, the dispersion relation
for the transverse permittivity contains the additional term
originating from the presence of a double pole at zero frequency,
whereas the longitudinal  permittivity satisfies the standard
dispersion relation valid for dielectric materials. We have also
obtained the dispersion relations expressing the dielectric
permittivities of graphene along the imaginary frequency axis via
their imaginary parts.

It was shown that the form of dispersion relations for the response
functions of graphene is unaffected by a presence of the branch
points whose position on the real frequency axis depends on the
magnitude of the wave vector. We emphasize that the dispersion
relations express the principle of causality and are valid for
any function which is analytic in the upper half-plane of complex
frequency. An application region of some function, satisfying the
dispersion relations, in the theoretical description of a definite
physical phenomenon is a different matter. However, the spatially
nonlocal dielectric permittivities of graphene considered above are
derived on the basis of first principles of quantum field theory in
the framework of the Dirac model using the polarization tensor.
Because of this, in the application region of this model, their
specific features, including the presence of a double pole at zero
frequency, are of doubtless physical significance.

As discussed in Sec.~I, the experimental data on measuring the
Casimir interaction in graphene systems are in good agreement with
theoretical predictions of the Lifshitz theory when describing the
electromagnetic response of graphene by means of the polarization
tensor \cite{20,21,22,23}. The Lifshitz theory using the polarization
tensor was also found in perfect agreement with the third law of
thermodynamics (the Nernst heat theorem) \cite{52,53}. However, the
predictions of the Lifshitz theory for metallic test bodies were
found in disagreement with the measurement data and with the Nernst
heat theorem when the response of metals at low frequencies is
described by the dissipative Drude model. An agreement is restored
when using the dissipationless plasma model at low frequencies where
it should not work.

The meaning of disagreement of the fundamental Lifshitz theory with
the measurement data should not be underestimated. Sometimes in the
literature the following formulations are used: ``experimental
measurements of the Casimir interaction between two metallic objects...
show a better agreement with the theoretical prediction using the
plasma model than with that of the Drude model" \cite{54} or ``somewhat
surprisingly, the less realistic dissipationless plasma model is in
better agreement with experiment than the Drude model" \cite{34}. In
several precision Casimir experiments, however, the Drude model was
excluded at the confidence level up to 99.9\% (see Refs.~\cite{24,25}
for a review). Moreover, in the differential force measurement, where
the theoretical predictions using the Drude and the plasma models differ
by up to a factor of 1000, the Drude model was conclusively excluded,
whereas the plasma model was shown to be in agreement with the
measurement data \cite{55}. Thus, the experimental situation
demonstrates not a better or worse agreement, but an exclusion of
the description by means of the Drude model and an agreement with the
description given by the plasma model.

Although a neglect by the dissipation of conduction electrons at low
frequencies cannot be considered as a satisfactory resolution of the
problem, one should, nevertheless, admit that the plasma model has
some important physical property which is missing in the Drude model.
The lesson of graphene suggests that this property is the double pole
at zero frequency which appears for graphene only at a nonzero wave
vector, i.e., only with account of the spatial dispersion. This
conclusion is in line with the spatially nonlocal phenomenological
permittivities of metals suggested in Refs.~\cite{43,44,45,46}, which
are almost coinciding with the Drude model for the on-the-mass-shell
fields but deviate from it for the fields off the mass shell and
possess the double pole at zero frequency.

Recently the experimental test for the response of metals to the
low-frequency s-polarized fields off the mass shell was suggested
\cite{56,57}. It is based on measuring the magnetic field of a
magnetic dipole oscillating in the proximity of metallic plate.
The point is that most of the experiments confirming the validity
of the Drude model were performed in the area of the propagating
waves on the mass shell. As to the area of the s-polarized
off-the-mass-shell waves, it remains little explored. Thus, the
available information for the surface plasmon polaritons is
restricted to only the area of p-polarized waves off the mass
shell \cite{58}. The total internal reflection technique makes it
possible to examine the response of metals to the off-the-mass-shell
fields, but for $k$ only slightly exceeding $\omega/c$ \cite{59,60,61}.
The methods used in the near field optical microscopy to surpass the
diffraction limit \cite{62,63} are also more suitable for the
p-polarized waves off the mass shell \cite{64}.

One can conclude that the already available information concerning
the response functions of graphene obtained on the solid basis of
quantum field theory should be used for a reanalysis of the
low-frequency electromagnetic response of metals in the area of
s-polarized waves off the mass shell where the necessary experimental
information is missing. In this respect, it seems prospective to
continue investigation of the 3D Dirac materials \cite{65} and to
generalize the obtained results for the case of more complicated
physical systems such as real metals.

\section*{ACKNOWLEDGMENTS}
The authors are grateful to M. Bordag (Leipzig University) for helpful discussions.
This work was partially funded by the Ministry of Science and Higher Education
of Russian Federation (``The World-Class Research Center: Advanced Digital
Technologies", contract No. 075-15-2022-311 dated April 20, 2022).

\appendix
\section{Involved integrals }
\setcounter{equation}{0}
\renewcommand{\theequation}{A\arabic{equation}}

Here, we calculate the integrals used in Sec.~IV. Thus, the integral which appears
in Eq.~({\ref{eq32}) is
\begin{equation}I_1=\int_{-b}^{b}\frac{dx\sqrt{b^2-x^2}}{x-\omega}.
\label{A1}
\end{equation}
\noindent
Using the result 1.2.53(9) in Ref.~\cite{51}, one can present this integral in the form
\begin{eqnarray}
&&
I_1=-\int_{-b}^{b}\frac{x dx}{\sqrt{b^2-x^2}}-\omega\int_{-b}^{b}\frac{dx}{\sqrt{b^2-x^2}}
\nonumber \\
&&~~~
+(b^2-\omega^2)\int_{-b}^{b}\frac{dx}{(x-\omega)\sqrt{b^2-x^2}}
\label{A2} \\
&&~~~
=-\pi\omega+(b^2-\omega^2)\int_{-b}^{b}\frac{dx}{(x-\omega)\sqrt{b^2-x^2}}.
\nonumber
\end{eqnarray}

The integral entering Eq.~(\ref{A2}) is rearranged to the form
\begin{eqnarray}
&&
\int_{-b}^{b}\frac{dx}{(x-\omega)\sqrt{b^2-x^2}}=
\int_{-b}^{0}\frac{dx}{(x-\omega)\sqrt{b^2-x^2}}
\nonumber \\
&&~~~~~~~~~~~~
+\int_{0}^{b}\frac{dx}{(x-\omega)\sqrt{b^2-x^2}}.
\label{A3}
\end{eqnarray}
\noindent
By changing the sign of the integration variable in the first integral on the
right-hand side of Eq.~(\ref{A3}), we easily obtain
\begin{equation}
\int_{-b}^{b}\frac{dx}{(x-\omega)\sqrt{b^2-x^2}}=-2\omega
\int_{0}^{b}\frac{dy}{(\omega^2-y^2)\sqrt{b^2-y^2}}.
\label{A4}
\end{equation}
\noindent
The last integral can be evaluated with the help of 1.2.50(10) in Ref.~\cite{51}
with the result
\begin{equation}
\int_{-b}^{b}\frac{dx}{(x-\omega)\sqrt{b^2-x^2}}=-2\omega\left\{
\begin{array}{ll}
0, & |\omega|<b, \\
\frac{\pi}{2|\omega|\sqrt{\omega^2-b^2}}, & |\omega|>b.
\end{array}
\right.
\label{A5}
\end{equation}

Substituting this result in Eq.~(\ref{A2}) and taking into account that
$\omega/|\omega|=1$ for $\omega>0$ and $\omega/|\omega|=-1$ for $\omega<0$,
we arrive at
\begin{equation}
I_1=\left\{
\begin{array}{ll}
-\pi\omega, & |\omega|<b, \\
-\pi\omega+\pi\sqrt{\omega^2-b^2}, & \omega>b, \\
-\pi\omega-\pi\sqrt{\omega^2-b^2}, & \omega<-b.
\end{array}\right.
\label{A6}
\end{equation}

The second integral which appears
in Eq.~({\ref{eq32})
\begin{equation}
I_2=\int_{-b}^{b}\!dx\frac{b-\sqrt{b^2-x^2}}{x^2}
\label{A7}
\end{equation}
\noindent
is a more simple one. By multiplying the numerator and denominator by
$b+\sqrt{b^2-x^2}$, one can rearrange it to the form
\begin{equation}
I_2=\int_{-b}^{b}\frac{dx}{b+\sqrt{b^2-x^2}}=
2\int_{0}^{b}\frac{dx}{b+\sqrt{b^2-x^2}}.
\label{A8}
\end{equation}

Introducing the integration variable $y=\sqrt{b^2-x^2}$, we obtain
\begin{eqnarray}
&&
I_2=2\int_{0}^{b}\frac{y dy}{(y+b)\sqrt{b^2-y^2}}
\label{A9} \\
&&~~
=2\int_{0}^{b}\frac{dy}{\sqrt{b^2-y^2}}-
2b\int_{0}^{b}\frac{dy}{(y+b)\sqrt{b^2-y^2}}
\nonumber \\
&&~~
=2\arcsin\left.\frac{y}{b}\right|_0^b+2\left.\sqrt{\frac{b-y}{b+y}}\right|_0^b
=\pi-2.
\nonumber
\end{eqnarray}

\section{Branch points}
\setcounter{equation}{0}
\renewcommand{\theequation}{B\arabic{equation}}

Here, we calculate the integrals of the form of $H^{\rm Tr}(\xi,k)$ and
$H^{\rm L}(\xi,k)$ in Eqs.~(\ref{eq52}) and (\ref{eq60}) along the semicircles
$C_{\rho}^{\,l}$ and $C_{\rho}^{\,r}$ around the branch points
$\omega=\mp v_Fk=\mp b$, respectively (see Fig.~1). We begin with
\begin{equation}
H_{C_{\rho}^{\,l}}^{\rm Tr}(\xi,k)=\int_{C_{\rho}^{\,l}}
\frac{\omega[\ve^{\rm Tr}\ok-1]}{\omega^2+\xi^2}\,d\omega.
\label{B1}
\end{equation}

The semicircle $C_{\rho}^{\,l}$ bypassed in the negative direction can be described
as $\omega=-b+\rho e^{i\varphi}$ where $\varphi$ varies from $\pi$ to 0.
The permittivity $\ve^{\rm Tr}\ok$ is given by the second and first lines  in
Eq.~(\ref{eq16}) when $\varphi$ varies from $\pi$ to $\pi/2$ and from $\pi/2$ and 0,
respectively. Substituting these expressions in Eq.~(\ref{B1}) and using
the equation of a semicircle, one finds
\begin{eqnarray}
&&
H_{C_{\rho}^{\,l}}^{\rm Tr}(\xi,k)=\frac{\pi\alpha kc}{2} \rho
\label{B2} \\
&&~\times
\left\{
\int_{\pi}^{\pi/2}d\varphi
\frac{e^{i\varphi}\sqrt{(-b+\rho e^{i\varphi})^2-b^2}}{(-b+\rho e^{i\varphi})
[(-b+\rho e^{i\varphi})^2+\xi^2]}\right.
\nonumber \\
&&~-i\left.
\int_{\pi/2}^{0}d\varphi
\frac{e^{i\varphi}\sqrt{b^2-(-b+\rho e^{i\varphi})^2}}{(-b+\rho e^{i\varphi})
[(-b+\rho e^{i\varphi})^2+\xi^2]}\right\}.
\nonumber
\end{eqnarray}

{}From this equation it is seen that
\begin{equation}
H_{C_{\rho}^{\,l}}^{\rm Tr}(\xi,k)\tR 0.
\label{B3}
\end{equation}

For the second branch point, we consider the integral
\begin{equation}
H_{C_{\rho}^{\,r}}^{\rm Tr}(\xi,k)=\int_{C_{\rho}^{\,r}}
\frac{\omega[\ve^{\rm Tr}\ok-1]}{\omega^2+\xi^2}\,d\omega,
\label{B4}
\end{equation}
\noindent
where the semicircle is described as
$\omega=b+\rho e^{i\varphi}$ and, again, $\varphi$ varies from $\pi$ to 0.
Here, however,
the permittivity $\ve^{\rm Tr}\ok$ is given by the first and second lines of
Eq.~(\ref{eq16}) when $\varphi$ varies from $\pi$ to $\pi/2$ and from $\pi/2$ and 0,
respectively. Substituting these expressions in Eq.~(\ref{B4}) and repeating the
same calculation as above using the equation of a semicircle, we obtain
\begin{equation}
H_{C_{\rho}^{\,r}}^{\rm Tr}(\xi,k)\tR 0,
\label{B5}
\end{equation}
\noindent
i.e., Eq.~(\ref{eq56}) in the main text is proven.

Now we consider the quantity
\begin{equation}
H_{C_{\rho}^{\,l}}^{\rm L}(\xi,k)=\int_{C_{\rho}^{\,l}}
\frac{\omega[\ve^{\rm L}\ok-1]}{\omega^2+\xi^2}\,d\omega
\label{B6}
\end{equation}
\noindent
related to the longitudinal permittivity of graphene. In this case the permittivity
is given by Eq.~(\ref{eq15}), i.e., it diverges at the branch points $\omega=\mp b$.
In the vicinity of the branch point $\omega=-b$ under consideration now,
$\ve^{\rm L}\ok$ is given by the second and first lines  in
Eq.~(\ref{eq15}) when $\varphi$ varies from $\pi$ to $\pi/2$ and from $\pi/2$ and 0,
respectively. Substituting these expressions in Eq.~(\ref{B6}) and using the equation of
 a semicircle $\omega=-b+\rho e^{i\varphi}$, we find
\begin{widetext}
\begin{eqnarray}
&&
H_{C_{\rho}^{\,l}}^{\rm L}(\xi,k)=\frac{\pi\alpha kc}{2}\left\{
\int_{\pi}^{\pi/2}d\varphi
\frac{\rho e^{i\varphi}(-b+\rho e^{i\varphi})}{\sqrt{(-b+\rho e^{i\varphi})^2-b^2}
[(-b+\rho e^{i\varphi})^2+\xi^2]}\right.
\nonumber \\
&&~
+i\left.
\int_{\pi/2}^{0}d\varphi
\frac{\rho e^{i\varphi}(-b+\rho e^{i\varphi})}{\sqrt{b^2-(-b+\rho e^{i\varphi})^2}
[(-b+\rho e^{i\varphi})^2+\xi^2]}\right\}.
\label{B7}
\end{eqnarray}
\end{widetext}

In the limiting case when $\rho$ goes to zero, Eq.~(\ref{B7}) reduces to
\begin{eqnarray}
&&
\lim_{\rho\to 0}H_{C_{\rho}^{\,l}}^{\rm L}(\xi,k)=
-\frac{\pi\alpha k c b}{2(b^2+\xi^2}
\label{B8} \\
&&~\times
\left\{
\int_{\pi}^{\pi/2}\! e^{i\varphi}\lim_{\rho\to 0}
\frac{\rho }{\sqrt{(-b+\rho e^{i\varphi})^2-b^2}}d\varphi\right.
\nonumber \\
&&~
+i\left.
\int_{\pi/2}^{0}\! e^{i\varphi}\lim_{\rho\to 0}
\frac{\rho }{\sqrt{b^2-(-b+\rho e^{i\varphi})^2}}d\varphi\right\}.
\nonumber
\end{eqnarray}

The limits under the sign of these integrals can be {\protect \\} easily calculated using the
l'H\^{o}pital's rule. For example,
\begin{equation}
\lim_{\rho\to 0}
\frac{\rho }{\sqrt{b^2-(-b+\rho e^{i\varphi})^2}}=
-\lim_{\rho\to 0}
\frac{\sqrt{b^2-(-b+\rho e^{i\varphi})^2}}{(-b+\rho e^{i\varphi})e^{i\varphi}}=0
\label{B9}
\end{equation}
\noindent
leading, due to Eq.~(\ref{B8}), to
\begin{equation}
\lim_{\rho\to 0}H_{C_{\rho}^{\,l}}^{\rm L}(\xi,k)=0.
\label{B10}
\end{equation}

The second branch point $\omega=v_Fk$ is considered in perfect analogy to the
above with the same result
\begin{equation}
\lim_{\rho\to 0}H_{C_{\rho}^{\,r}}^{\rm L}(\xi,k)=
\lim_{\rho\to 0}\int_{C_{\rho}^{\,r}}
\frac{\omega[\ve^{\rm L}\ok-1]}{\omega^2+\xi^2}\,d\omega =0.
\label{B11}
\end{equation}
\noindent
This concludes the proof of Eq.~(\ref{eq62})


\end{document}